\documentclass[aps,preprint]{revtex4}
\usepackage{bm,natbib}
\bibpunct{(}{)}{;}{a}{}{,}\gdef\harvardand{and}
\def\kB{k_{\rm B}}

\begin{document}

\title{On the stupendous beauty of closure}

\author{Hans Christian \"Ottinger}
\email[]{hco@mat.ethz.ch}
\homepage[]{http://www.polyphys.mat.ethz.ch/} \affiliation{ETH
Z\"urich, Department of Materials, Polymer Physics, HCI H 543,
CH-8093 Z\"urich, Switzerland}

\date{\today}

\begin{abstract}
Closure seems to be something rheologists would prefer to avoid. Here, the story of closure is told in such a way that one should enduringly forget any improper undertone of ``uncontrolled approximation'' or ``necessary evil'' which might arise, for example, in reducing a diffusion equation in configuration space to moment equations. In its widest sense, closure is associated with the search for self-contained levels of description on which time-evolution equations can be formulated in a closed, or autonomous, form. Proper closure requires the identification of the relevant structural variables participating in the dominant processes in a system of interest, and closure hence is synonymous with focusing on the essence of a problem and consequently with deep understanding. The derivation of closed equations may or may not be accompanied by the elimination of fast processes in favor of dissipation. As a general requirement, any closed set of evolution equations should be thermodynamically admissible. Thermodynamic admissibility comprises much more than the second law of thermodynamics, most notably, a clear separation of reversible and irreversible effects and a profound geometric structure of the reversible terms as a hallmark of reversibility. We discuss some implications of the intimate relationship between nonequilibrium thermodynamics and the principles of closure for rheology, and we illustrate the abstract ideas for the rod model of liquid crystal polymers, bead-spring models of dilute polymer solutions, and the reptation model of melts of entangled linear polymers.
\end{abstract}



\maketitle

\section{Introduction}
The term ``closure'' usually comes in the combination ``closure
approximation,'' and it is considered as a necessary evil in
deriving an autonomous, or closed, set of evolution equations in an
attempt to simplify or solve a problem. Employing a closure
approximation, usually of unclear and uncontrollable quality, then
appears as the desperate way out of a dead end.

The prototype scenario for closure approximations is the passage
from probability densities to moments. The evolution of a
probability density, or configurational distribution function, of a
Markov process is governed by some kind of Kolmogorov forward or
differential Chapman-Kolmogorov equation [\citet{Gardiner};
\citet{hcobook}], say a Fokker-Planck equation (or, a diffusion
equation in configuration space). The goal is to derive autonomous
evolution equations for a set of moments. The dilemma is that the
evolution equations for the moments following from the Fokker-Planck
equation involve more complicated moments than one intended to
consider. Adding these more complicated moments to the list of
variables does not help because their evolution involves even more
complicated moments. To avoid an intractable infinite hierarchy one
needs to approximate the more complicated moments in terms of the
simpler moments that one actually wants to work with. Only such a
closure approximation leads to an autonomous set of evolution
equations for a set of moments. In rheology, a number of systems are
treated by closure approximations of this type, such as
\emph{liquid-crystal polymers} [\citet{HinchLeal76}; \citet{Doi81};
\citet{DoiEdwards}; \citet{AdvaniTucker90}; \citet{Larson90};
\citet{hco52}; \citet{Bhaveetal93}; \citet{ChaubalLealFred95};
\citet{ChaubalLeal98}; \citet{FengChaubalLeal98}; \citet{hco124};
\citet{Edwards02}; \citet{ForestWang03}; \citet{KroegerAmmChin08}],
\emph{magnetic fluids} [\citet{Martsenyuketal74};
\citet{ZubarevIska00}; \citet{IlgKroger02}], or the nonlinear
effects of \emph{finitely extensible nonlinear elastic (FENE)
springs} [\citet{BirdDotsonJohnson80}; \citet{hco27}; \citet{hco31};
\citet{Wedgewoodetal91}; \citet{hco84}; \citet{Lielensetal99};
\citet{YuDuLiu05}; \citet{DuLiuYu05}; \citet{PrabhakarPra06}]
\emph{hydrodynamic interactions} [\citet{hco23}; \citet{hco33};
\citet{hco34}; \citet{hco35}; \citet{Wedgewood89}; \citet{hco77}]
and \emph{excluded volume} [\citet{hco116}; \citet{PrabhakarPra02};
\citet{Prakash02}] in dilute polymer solutions
[\citet{BirdetalDPLII}; \citet{hco57}]. The pioneer of closure for a
dumbbell model with a nonlinear spring force law was
\citet{Peterlin61,Peterlin66}. Experience based on a comparison of
the solutions of Fokker-Planck equations and moment equations
suggests that there seems to be enormous room for criticism and
improvement in the closure game.

A similar closure problem arises if one starts from a probability
density in a high-dimensional space and passes to a contracted
lower-dimensional distribution by integrating out degrees of
freedom. A classical challenge is the derivation of evolution
equations for the single- and two-particle distribution functions
from the Liouville equation for the probability density of a
macroscopically large number of particles. For pairwise
interactions, the equation for the $n$-particle distribution
function involves also the $(n+1)$-particle distribution function.
One thus arrives at the BBGKY hierarchy of equations for contracted
distribution functions introduced independently by Bogolyubov
(1946), Born and Green (1946), Kirkwood (1946), and Yvon (1937). In
this context, Boltzmann's celebrated \emph{Sto{\ss}zahlansatz}, in
which the two-particle distribution function at the beginning of a
collision is expressed as a product of single-particle distribution
functions, is the basis for the successful derivation of a closed
kinetic equation for the single-particle distribution. Also in this
context, closure has led to doubts and controversy, for example,
about its role in the emergence of irreversibility.

At this point, it should be quite obvious why closure is often
perceived as a necessary evil and as a questionable but unavoidable
mathematical trick with threateningly far-reaching consequences. Why
then would anybody associate stupendous beauty with closure, as
suggested in the title of this article? As illustrated above,
closure has to do with the search for simplified autonomous levels
of description, and \emph{proper simplification is the key to
understanding by focusing on the essence of a problem.} The search
for closure, understood as the physically motivated search for
simple autonomous levels of description for a given range of
phenomena of interest, is at the heart of developing insightful
theories leading to fundamental understanding and useful
applications in rheology and many other branches of science and
engineering. I see the beauty of closure in this association with
recognizing, highlighting and formulating the essentials, as I want
to elaborate in this article. I try to offer some colorful tesserae
which, hopefully, the reader can use to compose an appealing mosaic
of the role of closure in the process of understanding.

\section{The Beginning: Differential Constitutive Equations}
The origin of rheology as a scientific discipline lies in the 1920s
when Eugene C.~Bingham coined the term \emph{rheology} (1920) and
the \emph{Society of Rheology} was founded (1929). Closure, in a
wide sense, has been a central topic of rheology ever since the
derivation of the convected Maxwell model from molecular
considerations by \citet{GreenTobolsky46} and the famous formulation
of differential rheological equations of state by \citet{Oldroyd50}.
Differential constitutive equations for the stress tensor have a
long and successful history in rheology [see, for example,
\citet{BirdetalDPLI}]. Such constitutive equations successfully
describe a variety of nonlinear viscoelastic phenomena in a simple
setting, and hence they contribute significantly to our
understanding in rheology.

The flow behavior of Newtonian fluids can be described in terms of
the five hydrodynamic field of mass density, velocity (three
components), and temperature (instead of the velocity, one can use
the momentum density and, instead of the temperature, one can use
the internal energy or entropy density). This is a natural, universal
setting suggested by conservation laws. What is the minimum setting
for an autonomous description of more complex fluids? For
rheologists, it seems natural to employ the stress tensor as an
important further variable, and the search for differential
constitutive equations expresses the belief that, with the stress
tensor alone, one can obtain a closed description of complex fluids.
An additional \emph{tensor as a structural variable} provides a
classical setting for the autonomous description of complex fluids
in rheology.

While we rheologists grew up with differential rheological equations
of state, the possibility of formulating realistic equations of this
type is not \emph{a priori} obvious. The primary source of stresses
are forces acting over distances, and the stress tensor can be
expressed as the average of a tensor product of relative position
and force vectors [\citet{IrvingKirkwood50}]. A differential
equation of state implies the possibility of finding a closed
description for such an average. Describing complex fluids with just
an additional tensor always implies a closure assumption for an
average of some tensor product of two vectors. This becomes more
obvious when one tries to derive differential equations of state
from kinetic theory [\citet{BirdetalDPLII}] as, for example, in the
classical work by \citet{Peterlin61,Peterlin66} on nonlinear
springs. Kinetic theory also explains why it may be more natural and
convenient to work with a conformation tensor, such as the second
moment of a dumbbell vector, rather than with the stress tensor
directly. Of course, the stress tensor must then be expressed in
terms of the conformation tensor [\citet{hco115}].

In a wider sense, any choice of an autonomous level of description
admitting a closed description requires some form of closure
assumption. For example, when generalizing from dumbbell models of
polymer solutions to bead-spring-chain models one looks for closure
in a whole array of conformation tensors, where Gaussian
approximations are particularly popular and successful. Models with
two coupled tensor variables have also been motivated by abstract
orientation and elongation tensors [\citet{WilchFariaHutter01}] or
by the successful fitting of rheological data [\citet{EdwBerMav96}].
It is natural to include the heat flux in the same way as the
momentum flux into the list of variables. Working with a tensor and
a vector is hence another natural scenario [\citet{Mueller67};
\citet{BerisEdwards}; \citet{JouEIT}; \citet{Jouetalbib98};
\citet{JouCasasVaz01}; \citet{LebonEIT}] which is useful not only
for complex fluids undergoing non-isothermal flow
but also to formulate relativistic hydrodynamics
[\citet{hco109,hco111}]. The entire framework of \emph{extended
irreversible thermodynamics} has been built on the idea to employ
the momentum and heat fluxes as additional structural variables
[\citet{JouEIT}; \citet{Jouetalbib98}; \citet{JouCasasVaz01};
\citet{LebonEIT}]. The level of a tensor and a scalar has been
analyzed in its general form in \citet{hco143} because it has found
a number of modern applications, for example, in the Doi-Ohta model
of emulsions and blends [\citet{DoiOhta91}; \citet{hco108}], where
the scalar describes the amount of interface per unit volume and the
tensor variable separately accounts for the orientation of the
interface, or in the pompon model for melts of branched polymers
[\citet{McLeishLar98,hco130}], where the scalar represents the
stretch of a tube confining the molecular backbone and the tensor is
used to characterize the tube orientation, or in transient network
models of polymer melts, where the scalar describes the number
density of network segments and the tensor characterizes the segment
deformation. An even simpler example is provided by inhomogeneous
dilute polymer solutions [\citet{hco56}; \citet{BerisMavran94}],
where the scalar describes the polymer concentration and the tensor
represents the polymer stretch and orientation.

In all the above-mentioned theories with a tensor variable,
underlying closure assumptions must exist, at least, if one models
nonlinear effects. The success of differential equations of state
suggests not to consider closure as a necessary evil but as an
important step toward grasping the qualitative essence of rheological phenomena.
If one thinks about the Doi-Ohta or pompon models one appreciates
the understanding of the rheological behavior of complex systems
based on the evolution of structural variables, and one does not at
all worry about closure approximations. As a matter of fact, closure creates
comprehension.

\section{Thermodynamic Structure}
To understand the essence of rheological and other nonequilibrium
properties, one should formulate equations on as coarse as possible
levels of description. Passing to a coarser level of description
should not be regarded as a formal mathematical approximation but
rather as an insightful physical identification of the relevant
variables, that is, as an intellectual achievement. We are then
faced with the following question: \emph{What are the fundamental
physical principles that should be respected in formulating an
evolution equation on any autonomous level of description?} This
is, of course, the realm of nonequilibrium thermodynamics. In the
ideal case, the fundamental thermodynamic structure enforces not
only physically admissible but also mathematically well-behaved
equations for which the existence and uniqueness of solutions can be
demonstrated. The thermodynamic structure should also be preserved
in numerical integration schemes.

After respecting all conservation laws, the most prominent
thermodynamic principle is certainly the second law excluding the
possibility of a negative entropy production rate. Also the clear
separation of reversible and irreversible contributions to evolution
equations is of fundamental importance. The reversible contribution
should possess a rich structure reflecting the idea of ``mechanistic
control,'' and it should not touch the entropy. The reversible
contribution is generally assumed to be of the Hamiltonian form and
hence requires an underlying geometric structure (given in terms of
a Poisson bracket or Poisson operator). The remaining irreversible
contribution is driven by the nonequilibrium entropy by means of a
dissipative bracket. In the GENERIC (``general equation for the
nonequilibrium reversible-irreversible coupling'') framework of
nonequilibrium thermodynamics [\citet{hco99}; \citet{hco100};
\citet{hcobet}], these ideas are condensed into the evolution
equation
\begin{equation} \label{LMformulation}
  \frac{dx}{dt} = L \cdot \frac{\delta E}{\delta x} +
  M \cdot \frac{\delta S}{\delta x} ,
\end{equation}
where $x$ represents the set of independent variables required for a
complete description of a given nonequilibrium system, $E$ and $S$
are the total energy and entropy expressed in terms of the system
variables $x$, and $L$ and $M$ are certain linear operators, or
matrices, which can also depend on $x$. Equation
(\ref{LMformulation}) is supplemented by the complementary
degeneracy requirements
\begin{equation} \label{LSconsistency}
  L \cdot \frac{\delta S}{\delta x}=0 ,
\end{equation}
and
\begin{equation} \label{MEconsistency}
  M \cdot \frac{\delta E}{\delta x}=0 .
\end{equation}
The requirement that the entropy gradient $\delta S/\delta x$ is in
the null-space of $L$ in Eq.~(\ref{LSconsistency}) expresses the
reversible nature of the $L$-contribution to the dynamics: the
functional form of the entropy is such that it cannot be affected by
the operator generating the reversible dynamics. The requirement
that the energy gradient $\delta E/\delta x$ is in the null-space of
$M$ in Eq.~(\ref{MEconsistency}) expresses the conservation of the
total energy in a closed system by the $M$-contribution to the
dynamics. The two contributions to the time-evolution of $x$
generated by the energy $E$ and the entropy $S$ in
Eq.~(\ref{LMformulation}) are called the reversible and irreversible
contributions, respectively.

Further general properties of the matrices $L$ and $M$ are discussed
most conveniently in terms of the Poisson and dissipative brackets
\begin{equation} \label{Poissonbrackdef}
  \{A,B\} =  \frac{\delta A}{\delta x} \cdot L \cdot
  \frac{\delta B}{\delta x} ,
\end{equation}
\begin{equation} \label{dissipbrackdef}
  [A,B] = \frac{\delta A}{\delta x} \cdot M \cdot
  \frac{\delta B}{\delta x} ,
\end{equation}
where $A$, $B$ are sufficiently regular real-valued functions on the
space of independent variables. In terms of these brackets,
Eq.~(\ref{LMformulation}) and the chain rule lead to the following
time-evolution equation of an arbitrary function $A$ in terms of the
two separate generators $E$ and $S$,
\begin{equation} \label{brackform}
  \frac{dA}{dt} = \{A,E\} + [A,S] .
\end{equation}
The further conditions for $L$ can now be stated as the antisymmetry
property
\begin{equation} \label{condLasym}
  \{A,B\}=-\{B,A\} ,
\end{equation}
and the Jacobi identity
\begin{equation} \label{condLJacobi}
  \{A,\{B,C\}\}+\{B,\{C,A\}\}+\{C,\{A,B\}\}=0 ,
\end{equation}
whereas the product or Leibniz rule for Poisson brackets,
\begin{equation} \label{Leibnizrule}
  \{A B,C\} = A \{B,C\} + B \{A,C\} ,
\end{equation}
follows immediately from the definition in
Eq.~(\ref{Poissonbrackdef}). In these equations, $C$ is another
arbitrary sufficiently regular real-valued function on the state
space. These properties are well-known from the Poisson brackets of
classical mechanics, and they capture the essence of reversible
dynamics. The Jacobi identity (\ref{condLJacobi}), which is a highly
restrictive condition for formulating proper reversible dynamics,
expresses the invariance of Poisson brackets in the course of time
(\emph{time-structure invariance}).

Further properties of $M$ can be formulated in terms of the symmetry
condition
\begin{equation} \label{condMsym}
  [A,B]=[B,A] ,
\end{equation}
and the non-negativeness condition
\begin{equation} \label{condMpos}
  [A,A] \ge 0 .
\end{equation}
This non-negativeness condition, together with the degeneracy
requirement (\ref{LSconsistency}), guarantees that the entropy is a
nondecreasing function of time,
\begin{equation} \label{increntrop}
  \frac{dS}{dt} = \frac{\delta S}{\delta x} \cdot M \cdot
  \frac{\delta S}{\delta x} = [S,S] \ge 0 .
\end{equation}
The properties (\ref{condMsym}) and (\ref{condMpos}) imply the
symmetry and the positive-semidefiniteness of $M$ [for a more
sophisticated discussion of the \emph{Onsager-Casimir symmetry}
properties of $M$, see Sections 3.2.1 and 7.2.4 of \citet{hcobet}].
From a physical point of view, $M$ may be regarded as a friction
matrix.

The thermodynamic structure summarized here can also be regarded
as a \emph{geometric structure}. In the mathematical literature,
this geometric structure is sometimes referred to as
\emph{metriplectic} [\citet{Morrison86}]. In particular, Poisson
operators are intimately related to (duals of) Lie algebras [see,
for example, Appendix~B of \citet{hcobet} or
\citet{MarsdenRatiu}]. For example, convection effects are related
to the Lie group of space transformations and its representations;
as a result, we obtain convected rather than partial time
derivatives, as required by the famous principle of \emph{material
objectivity} or \emph{frame indifference} [\citet{BirdetalDPLI};
\citet{Lodge}].

\section{Doi and Bingham Closures}
We now have the thermodynamic tools to address the topic of closure thoroughly.
One of the most famous closures in rheology is the quadratic
\emph{ansatz}
\begin{equation}\label{Doiclo}
  - \dot{\bm{\gamma}} : \langle \bm{uuuu} \rangle =
  - \dot{\bm{\gamma}} : \langle \bm{uu} \rangle
  \langle \bm{uu} \rangle
\end{equation}
introduced by Doi in the theory of liquid crystal polymers
[\citet{Doi81}]. The Hess-Doi theory of liquid crystal polymers
[\citet{Hess76}, \citet{Doi81}] is based on a configurational
distribution function $f(\bm{u})$ for an ensemble of rigid rods,
where $\bm{u}$ is the orientation vector of a rod. The averages in
Eq.~(\ref{Doiclo}) are performed with the configurational
distribution function $f(\bm{u})$. If $\dot{\bm{\gamma}}$ is the sum
of the velocity gradient tensor and its transpose, the left-hand
side of Eq.~(\ref{Doiclo}) constitutes a convective contribution to
the time evolution of the second moment tensor $\langle \bm{uu}
\rangle$. As one wishes to obtain a closed evolution equation for
$\langle \bm{uu} \rangle$, the fourth moment has to be expressed in
terms of second moments, and Eq.~(\ref{Doiclo}) offers the simplest
possibility. An appealing alternative is to assume
\begin{equation}\label{Binghamclo}
  \langle \bm{uuuu} \rangle = \int \bm{uuuu} \,
  f_{\langle \bm{uu} \rangle}(\bm{u}) \, d^2u ,
\end{equation}
where $f_{\langle \bm{uu} \rangle}(\bm{u})$ is a given class of
distribution functions parametrized by the second moments
[\citet{ChaubalLeal98,FengChaubalLeal98}]. The most popular choice
is the exponential of a quadratic form of $\bm{u}$, the
mathematical-statistical properties of which have been studied by
C.~\citet{Bingham74} (who should not be confused with the pioneering
rheologist Eugene C.~Bingham).

The idea of introducing \emph{parametric distributions} to obtain
closure is widely used in nonequilibrium statistical thermodynamics.
Generalized canonical distribution functions parametrized by
Lagrange multipliers are obtained by maximizing the entropy under
constraints [\citet{hco147}; \citet{hco148}]. For example, the
Bingham distribution arises by maximizing the entropy for a fixed
second-moment tensor [\citet{hco148}]. The derivative of the entropy
with respect to the second-moment tensor is then given by $\kB
\bm{\Lambda}$, where $\bm{\Lambda}$ is the Lagrange multiplier
associated with $\langle \bm{uu} \rangle$ (the Bingham distribution
is proportional to $\exp \{ - \bm{\Lambda} : \bm{uu} \}$). Efficient
integration schemes and accuracy control, with the possibility of
changing the level of description upon a loss of accuracy, have been
proposed in the manifold of generalized canonical distribution
functions [\citet{hco147}; \citet{hco148}], which is also known as
the quasi-equilibrium manifold. The idea of parametric distributions
has been further developed into the powerful tool of the
\emph{invariant manifold method} [\citet{GorbanKarlin92};
\citet{GorbanKarlin94}; \citet{hco128}; \citet{GorbanKarlin}], where
geometric ideas and thermodynamic projectors are found to offer a
more elegant and general approach to closure problems than explicit
parametrizations, and an iterative Newton method conveniently
provides successive improvements of the equations for the moments.
However, the invariant manifold method cannot produce any
dissipation; in particular, it cannot produce irreversible equations
from reversible ones. To allow for this possibility, the method has
been enhanced by \emph{Ehrenfest coarse graining} [\citet{hco128};
\citet{GorbanKarlin}].

The closure approximations (\ref{Doiclo}) and (\ref{Binghamclo}),
among many others, have been studied intensely in the literature on
the rod model of liquid crystal polymers [\citet{HinchLeal76};
\citet{Doi81}; \citet{DoiEdwards}; \citet{AdvaniTucker90};
\citet{Larson90}; \citet{hco52}; \citet{Bhaveetal93};
\citet{ChaubalLealFred95}; \citet{ChaubalLeal98};
\citet{FengChaubalLeal98}; \citet{hco124}; \citet{Edwards02};
\citet{ForestWang03}; \citet{KroegerAmmChin08}]. Some of these
closures are extremely successful in particular flow situations but
fail terribly in others. For example, in simple shear flow, Doi's
quadratic closure (\ref{Doiclo}), when applied to the fourth moments
occurring both in the reversible and in the irreversible terms,
exhibits only time-independent stable solutions
[\citet{Bhaveetal93}; \citet{ChaubalLealFred95}] and thus misses the
well-known periodic solutions known as ``tumbling'' (preferred axis
of alignment rotates in the plane of shear) and ``wagging''
(preferred axis oscillates back and forth in the plane of shear),
whereas the Bingham closure admits such time-dependent stable
solutions. Can one identify the ultimate winners and losers by such
observations so easily?

At this point, we should remember our goals and ambitions. We are
not really interested in the Olympic Games of mathematical closure
approximations, in which gold, silver, and bronze medals are given
away in a myriad of specialized flow disciplines. We rather strive
after fundamental understanding by identifying and verifying
autonomous levels of description bringing out the essence of a
problem, without necessarily reproducing all the details. In the
present case, the autonomous level of description for liquid crystal
polymers is proposed to be given by the structural variable $\langle
\bm{uu} \rangle$, and we must respect the structure of
thermodynamically admissible equations on this level. Moreover,
judging the success of closure approximations by a comparison with
the exact results for all kinds of flow situations is not
particularly useful; we should clearly prefer to find \emph{a
priori} criteria for overall success that do not require any
knowledge of exact solutions.

As the fourth moment in Eq.~(\ref{Doiclo}) is associated with
convection, we deal with a reversible term generated by the energy
gradient with the help of a Poisson operator. As the kinetic energy,
which generates convection, is unaffected by the structural
variable, we need to focus on the Poisson operator and its
properties as the hallmarks of reversible motion under mechanistic
control. The most restrictive criterion is the Jacobi identity
(\ref{condLJacobi}) expressing the time-structure invariance of
reversible dynamics. This criterion has been analyzed in great
detail in a similar context by \citet{hco102}, and we here merely
summarize and discuss the most important results.

According to \citet{hco102}, reversible dynamics obtained under the
Doi closure (\ref{Doiclo}) can be generated by a valid Poisson
operator, whereas this is impossible for the Bingham closure
(\ref{Binghamclo}) and many other common closures. The analysis of
\citet{hco102} proceeds according to symmetry. Note that the fourth
moment is symmetric in all four tensor indices. This full symmetry,
which is respected by the Bingham closure, is incompatible with
time-structure invariance. The Doi closure possesses symmetry in
pairs of indices, and under exchanging the pairs. For this lowered
level of symmetry, the Doi closure (\ref{Doiclo}) is the only
possible closure compatible with time-structure invariance. For even
lower levels of symmetry, further admissible closures can be
constructed [\citet{hco102}].

Of course, you could say ``Why should I, as a rheologist, make my
life more complicated by worrying about some mysterious Jacobi
identity?'' Because Poisson structures are at the heart of
reversible dynamics, and because violating laws out of ignorance
does not protect you from punishment! Of course, one could still try
to argue why a law is not applicable in a particular situation, but
the only reason for losing mechanistic control in reversible
dynamics that I am currently aware of is the presence of
nonholonomic constraints.

If one has to \emph{choose or compromise between symmetry and
time-structure invariance}, why should symmetry be less important?
The answer should be clear by now: Because our goal is to establish
a healthy autonomous level of description based on second moments,
consistent with all the laws of nonequilibrium thermodynamics, and
not necessarily to achieve a faithful mathematical approximation of
a fourth moment in terms of second moments.

If only the quadratic closure (\ref{Doiclo}) is thermodynamically
admissible, does thermodynamics thus force us into an inferior
closure that cannot even predict ``tumbling'' and ``wagging'' in
shear flow? A much more balanced view was offered by
\citet{Bhaveetal93}, \citet{ChaubalLealFred95},
\citet{ChaubalLeal98}, and \citet{FengChaubalLeal98}, based on a
``solution map'' or ``bifurcation set'' in the parameter space of
two-dimensional flows and nematic strength. \emph{Shear flow appears
as a singular special case}, and ``tumbling'' and ``wagging'' do
actually occur within the quadratic closure for flows that are only
very slightly more rotational than simple shear flow. The overall
solution map is deformed only slightly for the quadratic closure,
but with rather serious consequences for the singular special case
of shear flow. Moreover, none of the known closures exhibiting
``tumbling'' and ``wagging'' can predict the proper transition from
``wagging'' to flow aligned steady solutions at high shear rates
[\citet{FengChaubalLeal98}].

A more rotational behavior with ``tumbling'' and ``wagging'' can
also be achieved by going from the upper convected codeformational derivatives
appearing naturally in the second moment equations to mixed or
Schowalter derivatives [see pp.~556 and 568 of \citet{BerisEdwards}].
It has been elaborated in Section 4.2.3 of \citet{hcobet} that,
within the GENERIC framework, such a modification can be
implemented through an additional irreversible contribution, as
suggested by the occurrence of a slip coefficient. Schowalter
derivatives are represented by an antisymmetric contribution to the
friction matrix so that, according to Eq.~(\ref{increntrop}), they
do not lead to entropy production.

We have focused entirely on the fourth moment (\ref{Doiclo}) that
appears as a convective contribution to the time evolution of the
second moment tensor $\langle \bm{uu} \rangle$. We have not paid any
attention to the fact that, for the Maier-Saupe mean-field nematic
potential, there occurs another fourth moment in the irreversible
contribution to that evolution equation so that a further term
requires closure. As a matter of fact, a different closure could be
used there because irreversible contributions are much less
restricted than reversible ones. For example, one could use the Doi
closure in the reversible contribution to fulfill time-structure
invariance, and the more symmetric Bingham closure in the
irreversible contribution in order to combine the advantages of both
closures [\citet{SgalariLealFeng95}], where the less restricted
closure in the irreversible term seems to be the bigger source of
problems. Actually, the fact that the GENERIC framework of
thermodynamics expresses time evolution in terms of the generators
$E$, $S$ and the matrices $L$, $M$ strongly suggests to consider
each of these building blocks separately, that is, with the natural
possibility of \emph{separate closures in the reversible and
irreversible terms}. In the subsequent section, we look at the
implications of thermodynamics for closure in the irreversible
contribution to the time evolution of moments. We do this in the
context of the Gaussian approximation, which has found many
successful applications in polymer kinetic theory. The rod model of
liquid crystal polymers could be investigated in a similar way, but
the handling of Gaussian distributions is more familiar.

\section{Gaussian Closure}
In the Gaussian closure procedure, or Gaussian approximation, the
Jacobi identity is not an issue. Because there are no constraints
and Gaussian objects are deformable, one can realize codeformational
behavior in the form of upper convected derivatives, which are known
to be consistent with a Poisson bracket [\citet{hcobet}].
Nevertheless, something remains to be checked for the irreversible
term. As Gaussian approximations are usually implemented on the
level of time-evolution equations, one needs to check whether they
are consistent with a Gaussian entropy generating irreversible
dynamics via the friction matrix obtained for Gaussian distributions
according to Eq.~(\ref{LMformulation}).

To discuss the proper formulation of the irreversible term in the
Gaussian approximation, we start from a Fokker-Planck equation of
the general form
\begin{equation}\label{FPEgen}
   \frac{\partial f}{\partial t} =
   - \frac{\partial}{\partial x_j} \left( A_j -
   \frac{1}{2} D_{jk} F_k \right) f
   + \frac{1}{2} \frac{\partial}{\partial x_j} D_{jk}
   \frac{\partial}{\partial x_k} f
\end{equation}
where $f=f(x)$ is a probability density in some $K$-dimensional
space with coordinates $x_j$, $A_j$ is a reversible drift vector,
the positive-semidefinite symmetric diffusion matrix $D_{jk}$
describes the irreversible effects,
\begin{equation}\label{forcedef}
   F_j = - \frac{\partial \ln f^{\rm eq}}{\partial x_j}
\end{equation}
is the effective force implied by the equilibrium probability
density $f^{\rm eq}$, and Einstein's summation convention is assumed
(one needs to sum from $1$ to $K$ over all indices occurring twice). We
look for (because we believe in the adequacy of) a description on
the level of a matrix $c$ of second moments with entries,
\begin{equation}\label{secmomdef}
   c_{ij} = \langle x_i x_j \rangle = \int x_i x_j f d^K x ,
\end{equation}
where the coordinates $x_j$ are assumed to be introduced such that
$c^{\rm eq}_{ij} = \delta_{ij}$. We further assume that the first
moments vanish for symmetry reasons. In the Gaussian approximation,
these assumptions imply $F_j = x_j$.

The entropy on the level of second moments can be obtained by
evaluating the Boltzmann-type conformational entropy $-\kB \int f
\ln (f/f^{\rm eq}) d^Kx$, where $\kB$ is Boltzmann's constant, for
Gaussian distributions. The result is [see, for example, Exercise 66
of \citet{hcobet}]:
\begin{equation}\label{Gaussentropy}
    S = \frac{1}{2} \kB \left( K - c_{jj} + \ln\det c \right) ,
\end{equation}
with the derivative
\begin{equation}\label{Gaussentropyder}
    \frac{\partial S}{\partial c_{ij}} = \frac{1}{2}
    \kB \left( c^{-1}_{ij} - \delta_{ij} \right) .
\end{equation}
Equation (\ref{Gaussentropy}) for the entropy is an essential
feature of any Gaussian approximation.

For the explicit transformation rule (\ref{secmomdef}) from
probability densities $f$ to second moments $\langle x_i x_j
\rangle$, one can transform the friction matrix $M$ occurring in the
irreversible contribution to the Fokker-Planck equation
(\ref{FPEgen}) according to Eq.~(6.180) of \citet{hcobet}. The
result is:
\begin{equation}\label{MGauss}
   2 \kB M_{ij,kl} = \langle x_i x_k D_{jl} \rangle
   + \langle x_i x_l D_{jk} \rangle
   + \langle x_j x_k D_{il} \rangle
   + \langle x_j x_l D_{ik} \rangle .
\end{equation}
All four contributions on the right-hand side of this equation are
equivalent if $M_{ij,kl}$ is contracted with symmetric tensors;
otherwise, the four contributions imply that the contraction should
be done with symmetrized tensors only. The Gaussian approximation
can now be introduced into Eq.~(\ref{MGauss}) by using Wick's
theorem to reduce the order of the moments (see, for example,
Eq.~(2.61) of \citet{hcobook}). We successively find
\begin{equation}\label{MGauss1}
   \langle x_i x_k D_{jl} \rangle = c_{ik} \langle D_{jl} \rangle
   + c_{mk} \left\langle x_i \frac{\partial D_{jl}}{\partial x_m}
   \right\rangle ,
\end{equation}
and
\begin{equation}\label{MGauss2}
   \langle x_i x_k D_{jl} \rangle = c_{ik} \langle D_{jl} \rangle
   + c_{mi} c_{nk} \left\langle \frac{\partial^2 D_{jl}}{\partial x_m
   \partial x_n} \right\rangle .
\end{equation}
The first term on the right-hand side of each of the
Eqs.~(\ref{MGauss1}) and (\ref{MGauss2}) represents the effect of
a self-consistently averaged diffusion matrix, whereas the second
term accounts for fluctuation effects. The description of
fluctuation effects by second-order derivatives looks particularly
natural. Equation (\ref{MGauss1}) can be rewritten in the
alternative form
\begin{equation}\label{MGauss3}
   c^{-1}_{kl} \langle x_i x_k D_{jl} \rangle = \langle D_{ij} \rangle
   + \left\langle x_i \frac{\partial D_{jk}}{\partial x_k}
   \right\rangle .
\end{equation}

As we have found the natural entropy and friction matrix for any
Gaussian closure, we can now compare to the evolution of the
second moments obtained directly from the Fokker-Planck equation
(\ref{FPEgen}),
\begin{equation}\label{FPEgenmom}
   \frac{\partial c_{ij}}{\partial t} = \langle A_i x_j \rangle
   + \langle x_i A_j \rangle
   - \frac{1}{2} \langle D_{ik} F_k x_j \rangle
   - \frac{1}{2} \langle x_i F_k D_{kj} \rangle
   + \langle D_{ij} \rangle
   + \frac{1}{2} \left\langle x_i
   \frac{\partial D_{jk}}{\partial x_k} \right\rangle
   + \frac{1}{2} \left\langle x_j
   \frac{\partial D_{ik}}{\partial x_k} \right\rangle .
\end{equation}
With Wick's theorem in the form of Eq.~(\ref{MGauss3}) and the
Gaussian property $F_j = x_j$, Eq.~(\ref{FPEgenmom}) can be
rewritten as
\begin{equation}\label{FPEgenmoms}
   \frac{\partial c_{ij}}{\partial t} = \langle A_i x_j \rangle
   + \langle x_i A_j \rangle
   + \frac{1}{2} ( \langle x_i x_k D_{jl} \rangle
   + \langle x_j x_k D_{il} \rangle )
   \left( c^{-1}_{kl} - \delta_{kl} \right) .
\end{equation}
The irreversible contribution to this evolution equation is exactly
what one recovers from the irreversible contribution to GENERIC by
combining Eqs.~(\ref{Gaussentropyder}) and (\ref{MGauss}). The
Gaussian approximation is thus nicely consistent with an
irreversible contribution to dynamics generated by the Gaussian
entropy on the level of second moments. Contrary to our
disappointing experience with the reversible term, parametric
density estimation works very nicely for Gaussian approximations to
the irreversible term.

In the Gaussian approximation, a configurational distribution
function $f$ is assumed to be Gaussian at any time $t$. This does
not necessarily imply a Gaussian stochastic process, for which all
joint distributions at different times must also be Gaussian. The
construction of a full Gaussian process governed by a linear
stochastic differential equation has been described and discussed
critically in Section 4.2.4 of \citet{hcobook} and by
\citet{hco150}. A well-defined and physically consistent stochastic
process on the level of second moments can be introduced by adding
noise to Eq.~(\ref{FPEgenmoms}) according to the
fluctuation-dissipation theorem [\citet{hco150}].

Because the Gaussian approximation has been used very successfully
in the \emph{kinetic theory of dilute polymer solutions}, where it
has been applied to hydrodynamic interactions [\citet{hco33};
\citet{hco34}; \citet{hco35}; \citet{Wedgewood89}; \citet{hco77}],
excluded volume [\citet{hco116}; \citet{PrabhakarPra02};
\citet{Prakash02}], and internal viscosity [\citet{Schieber93}], we
here specialize our general results to the case of bead-spring
chains, which has previously been considered by \citet{hco150}. If
the chains consist of $N$ beads, we have $K = 3(N-1)$
configurational degrees of freedom and the matrix $c$ is an array of
$(N-1) \times (N-1)$ tensors $\bm{c}_{jk}$ representing the
variances and covariances of the $N-1$ connector vectors, where the
normalization condition at equilibrium now reads $\bm{c}^{\rm
eq}_{jk} = \delta_{jk} \bm{1}$. According to
Eq.~(\ref{Gaussentropy}), the conformational entropy per polymer
molecule is given by
\begin{equation}\label{Gaussentropybsc}
    s_{\rm p} = \frac{1}{2} \kB \left[ \sum_{j=1}^{N-1} {\rm tr}
    ( \bm{1} - \bm{c}_{jj} ) + \ln\det c \right] ,
\end{equation}
where $c$ is the large matrix consisting of $(N-1) \times (N-1)$
blocks $\bm{c}_{jk}$, each of which is represented by a $3 \times
3$ matrix. As the GENERIC framework involves only derivatives of
entropy, we provide the result (\ref{Gaussentropyder}) in the form
\begin{equation}\label{Gaussentropyderbsc}
    \frac{\partial s_{\rm p}}{\partial \bm{c}_{jk}} = \frac{1}{2}
    \kB \left( \bm{c}^{\rm I}_{jk} - \delta_{jk} \bm{1} \right) ,
\end{equation}
where the tensors $\bm{c}^{\rm I}_{jk}$ are represented by the
blocks of the inverse of the large matrix $c$,
\begin{equation}\label{cIdef}
    \sum_{l=1}^{N-1} \bm{c}_{jl} \cdot \bm{c}^{\rm I}_{lk} =
    \delta_{jk} \bm{1} .
\end{equation}

The degeneracy condition (\ref{LSconsistency}) of nonequilibrium
thermodynamics implies that the entropic spring contribution to the
pressure tensor is given by
\begin{equation}\label{presstens}
    \bm{\Pi} = n_{\rm p} T \left\{ s_{\rm p} \bm{1}
    + \sum_{j,k=1}^{N-1} \left[ \bm{c}_{jk} \cdot \left(
    \frac{\partial s_{\rm p}}{\partial \bm{c}_{jk}} \right)^T
    + \left( \bm{c}_{jk} \right)^T \cdot
    \frac{\partial s_{\rm p}}{\partial \bm{c}_{jk}} \right]
    \right\} ,
\end{equation}
where $T$ is the absolute temperature and $n_{\rm p}$ the number
density of polymers. By inserting the derivatives
(\ref{Gaussentropyderbsc}) into Eq.~(\ref{presstens}) and using
the symmetry of the large matrix $c$ (which implies $( \bm{c}_{jk}
)^T = \bm{c}_{kj}$ and an analogous identity for the inverse), we
obtain
\begin{equation}\label{presstensc}
    \bm{\Pi} = n_{\rm p} s_{\rm p} T \, \bm{1}
    + n_{\rm p} \kB T \sum_{j=1}^{N-1} ( \bm{1} - \bm{c}_{jj} ) .
\end{equation}
Note that this simple form of the pressure tensor consisting of
Hookean spring contributions from each of the connectors is a
direct consequence of the functional form of the entropy in
Eq.~(\ref{Gaussentropybsc}).

\section{Variables Are Everything}
For the development and discussion of the Doi, Bingham, and Gaussian
closures in the preceding sections, we looked in great detail at
complicated moments and their expression in terms of second moment
tensors. Over all these details we should not forget that our sole
goal was to establish an autonomous level of description based on
second moment tensors. The important questions are: Can this be
done? How can this be done? An important outcome of the general
analysis of the second-moment level of description is: There is not
much choice in the reversible dynamics.

The key problem of nonequilibrium thermodynamics or, if you like, of
coarse graining, or of closure, or of understanding, is the choice
of good variables for a problem of interest. For this crucial task
of choosing variables, thermodynamics leaves you alone with your
insight, intuition, imagination, and ingenuity (${\rm i}^4$). To
gain insight, of course, experimental results are
of particular importance. Once
you have chosen your variables and expressed them in terms of the
variables of a more detailed well-established level of description,
often the atomistic level, \emph{statistical thermodynamics provides
systematic recipes for calculating the thermodynamic building
blocks} $E$, $S$, $L$, and $M$ [\citet{hco101}; \citet{hco131};
\citet{hcobet}; \citet{hco173}; \citet{hco182}]. These building
blocks imply the autonomous time evolution (\ref{LMformulation}) and
hence all closure properties. The generalized microcanonical,
canonical, and mixed ensembles of nonequilibrium statistical
mechanics can be considered as natural candidates for parametric
density estimation. Any concrete realization of the
GENERIC structure provides a solution to the closure problem.

The importance of the choice of variables can be illustrated nicely
for the reptation model of melts of entangled linear polymers. One
usually does not speak about closure in this context, however, one
clearly uses mean-field type and further simplifying assumptions.
Starting from the picture of a large number of massively entangled
random walk chains, one first assumes that one can look at the
motion of a single probe chain constrained by a tube or slip links
produced by other chains, or anisotropic friction accounting for the
hindrance of sideway motions by other chains. In a second step, the
single-chain picture is further reduced to that of a single segment
with orientation $\bm{u}$ at a position $s$ within the chain, where
the label $s$ varies from $0$ to $1$ in going from one chain end to
the other. In order to achieve this further simplification, one
needs to make an assumption like ``independent alignment of large
straight segments'' [\citet{DoiEdwards78a}, \citet{DoiEdwards78b},
\citet{DoiEdwards78c}] or ``smooth curvature''
[\citet{CurtissBird81a}, \citet{CurtissBird81b}]. The entanglement
length scale is thus introduced as the length of the independently
aligned segments or as the persistence length of smoothly curved
chains. The truly heroic idea of the reptation model is to postulate
that the single-segment configurational distribution function
$f(\bm{u},s)$ leads to an autonomous level of description for the
complicated system of entangled chains. This postulate implies a
natural but highly nontrivial closure.

The formulation of the thermodynamic building blocks, and hence of
the evolution equations, is a much simpler step than the choice of
variables and can actually be done with very little of ${\rm i}^4$,
as has been shown by the author in Section 8.4.6 of \citet{hcobet}.
Once $f(\bm{u},s)$ has been identified as a good structural
variable, there is so much guidance from nonequilibrium
thermodynamics that, for example, one is automatically reminded to
consider constraint release associated with the reptation of
constraining chains and the possibility of anisotropic tube cross
sections.

\section{Reduction Versus Coarse Graining}
For diffusion equations with Gaussian solutions, which are
associated with linear stochastic differential equations, the
second-moment equations reproduce all the features of the exact
solutions. Closure is usually considered as an approach to achieve
approximate solutions to nonlinear problems, with the ambition to be
as accurate as possible. This is the idea of a \emph{solution or
reduction technique}. In the context of the invariant manifold
method, we have already encountered the \emph{hallmark of reduction
techniques}: they do not produce any additional dissipation and, in
particular, reduction techniques cannot lead from reversible to
irreversible equations. Equation (\ref{MGauss}) for the Gaussian
approximation also exhibits the hallmark of reduction: the friction
matrix on the level of moments is directly proportional to the
diffusion tensor $D_{jk}$ on the level of the configurational
distribution function. Also dynamic renormalization, as carried out
in the context of hydrodynamic interactions in dilute polymer
solutions [\citet{hco185}], turns out to be a reduction technique.

To illustrate the closure problem in the introduction, we had also
mentioned the BBGKY hierarchy and Boltzmann's kinetic equation,
where the former is reversible and the latter is irreversible.
Boltzmann's derivation of his irreversible kinetic equation for
rarefied gases from the reversible equations of classical mechanics
is an enormous achievement that has created a lot of controversy and
deep insights [\citet{Lanford75ip}]. The derivation of Boltzmann's
kinetic equation cannot be achieved by a reduction technique. When
new dissipative processes arise in the passage from a more detailed
to a less detailed level of description, we speak of a
\emph{coarse-graining technique}.

In the case of Boltzmann's kinetic equation, the clear separation of
two time scales, namely the duration of collisions and the time
between collisions, is at the origin of irreversibility. A detailed
analysis of two-particle collisions is required to obtain the
Boltzmann equation which itself cannot resolve any processes on the
short scale of the duration of a collision. In general, in coarse
graining one treats fast processes as fluctuations and fluctuations
are associated with dissipation. \emph{Projection operators} provide
a powerful tool to separate fast and slow processes, thus providing
the statistical mechanics of coarse graining [\citet{Zwanzig61};
\citet{Mori65}; \citet{Mori65a}; \citet{Robertson66};
\citet{Grabert}; \citet{hcobet}].

Note that also the diffusion equations for configurational
distribution functions used in polymer kinetic theory themselves are the result
of coarse graining. In particular, spring forces between reference
points in linear polymer molecules have been discussed by
\citet{UnderhillDoyle04} by means of statistical mechanics in constant
extension and constant force ensembles. If, however, one is
interested in spring forces between beads, each located at the
center of mass of the smaller units coarse grained into the bead, the
force laws are different [\citet{hco51}, \citet{hco185}].

Both reduction and coarse graining lead to autonomous evolution
equations as a result of passing from a more to a less detailed
level of description. Both techniques must preserve the GENERIC
structure of nonequilibrium thermodynamics, both techniques are
associated with closure. Whereas the approximate character of
closure may be viewed as a disadvantage in reduction, it is the
actual goal of coarse graining: fast processes are eliminated in
favor of dissipative processes. This is understanding by focusing on
the relevant slow features. One should always try to obtain more
transparent equations by coarse graining and only when further
coarse graining is impossible, reduction and explicit solution
techniques should be used. With the tremendous increase of computer
power, there is the temptation of producing numerical solutions
without exploiting the full potential of coarse graining. Giving in
to this temptation, we miss the opportunity of deeper understanding
and, in the long run, our problem-solving skills will deteriorate.

\section{Rheology Without Closure}
We have recognized the closure problem, in a wide sense, as the
search for autonomous levels of description on which a closed
formulation of the evolution of systems is possible. An autonomous
system remembers its history, in particular, its flow history, only
through the current state of the structural variables of the level
of description. There is no explicit memory and the GENERIC
evolution equations (\ref{LMformulation}) are pure first-order
differential equations.

In rheology, however, also integral rheological constitute equations
and memory integral expansions are well-known
[\citet{BirdetalDPLI}]. In view of the enormous possibilities of
formulating memory functionals, it is very difficult both to select
an appropriate memory functional for a given problem and to
formulate the proper thermodynamic principles for memory functionals
in general [\citet{ColemanNoll63}]. In particular, there is no
direct scheme for calculating memory functionals from atomistic
models by means of statistical mechanics. Whereas some established
classes of integral models can certainly describe rheological
measurements for a variety of systems qualitatively well, and there
usually are sufficiently many fitting parameters to get quantitative
agreement, there are no systematic ways of translating molecular
understanding of a system into successful memory integral models.
The identification of key processes and relevant variables,
supported by statistical mechanics in evaluating thermodynamic
building blocks, is a much more promising pathway to understanding.

Of course, there are connections between differential and integral
rheological equations of state or, more generally, between
autonomous levels of description and memory functionals. By solving
or integrating the evolution equations for the structural variables,
in principle, one can determine memory functionals. However,
explicit solutions can only be obtained in exceptional (linear)
cases. For example, for the reptation model, a deformation measure
and a memory function can be identified such that its solution can
be written as a simple memory integral; the reptation model of Doi
and Edwards has actually been recognized as an integral model of the
K-BKZ type [see, for example, \citet{Hassager81}].

Another famous class of evolution equations with memory are
\emph{mode coupling theories}. As the memory effects are included by
following well-defined recipes [\citet{Gotze84}], one might like to
speak of mode coupling as ``quasi-closure.'' In particular, mode
coupling theory has been found to be useful in the discussion of the
glass transition [\citet{Gotze84}; \citet{KobAndersen94};
\citet{KobAndersen95}]. The glass transition problem is considered
to be a very difficult one because the ingenious structural variable
for describing it by autonomous evolution equations has not yet been
discovered. The potential energy landscape and the inherent
structure energy are clearly important concepts
[\citet{StillingerWeber82}; \citet{Angell95}; \citet{Stillinger95};
\citet{LaNaveScioretal03}], and the inherent structure pair
correlation function has been proposed as a promising candidate for
a coarse grained description of the approach to the glass transition
[\citet{hco167}]. This idea has actually revealed a static signature
of the onset of the glass transition [\citet{hco177}]. Structural
variables associated with ``shear transformation zones'' may be an
interesting alternative [\citet{FalkLanger98}].

\section{Closure}
We have offered some reflections on the topic of closure which
strongly suggest that closure should not be considered as a helpful
mathematical approximation but as the cornerstone of establishing
autonomous levels of description. Choosing good structural variables
is the physical prerequisite to achieve closure and to perform
successful coarse graining, that is, to elaborate the essential
features of problems. Closed simplified descriptions lead to
understanding. We have illustrated this situation in the context of
the highly restricted fourth-moment closure in the convection
term of the rod model of liquid crystal polymers, of the
admissibility of Gaussian closures in the irreversible term for
bead-spring models of dilute polymer solutions, and of the
far-reaching ingenuity of the variables in the reptation model of
melts of entangled linear polymers. Liquid crystals and Gaussian
closures provide examples of reduction,
that is, approximate solution procedures. Even more revealing is
closure obtained from a coarse graining procedure, which replaces
fast processes by dissipative ones, such as the reptation process.

With mild exaggeration (rooted in enthusiasm rather than
one-sidedness), our reflections on closure can be expanded into a
full world-map of rheology. The milestones of theoretical rheology
are ground-breaking new models, and these come with and actually
\emph{via} the insightful and revealing discovery of good variables.
Then, closed equations can be formulated on the corresponding
autonomous level of description. Nonequilibrium thermodynamics
provides the structure of admissible evolution equations,
nonequilibrium statistical mechanics offers the recipes for finding
concrete realizations of the thermodynamic structure for given
variables. We thus obtain all rheological information for the system
defined through the ingeniously chosen variables.

Internal energy and entropy are key concepts in nonequilibrium
thermodynamics. This clearly indicates the way to go for rheology.
Focusing on the momentum balance and momentum flux has been a good
starting point, but a detailed investigation of the associated
entropy flux must be the next step. This also requires consideration
of compressibility effects. How do we need to generalize the
concepts of thermal conductivity and heat capacity in complex fluids
undergoing flow? Pioneering experiments on anisotropic heat flow
have been performed by Venerus, Schieber and coworkers
[\citet{Venerusetal99}; \citet{Broermanetal99}; \citet{Iddiretal00};
\citet{Venerusetal04}; \citet{Schieberetal04};
\citet{Balasubetal05}], and some issues associated with generalizing
the concept of heat capacity have been discussed by \citet{hco183}.

Nonequilibrium thermodynamics guides the formulation of autonomous
evolution equations and offers the recipes for finding the
thermodynamic building blocks on a less detailed level of
description from a more detailed level of description. The
thermodynamic structure is a geometric one and the passage between
different levels of description hence is a mathematical topic of
structure preserving transformations [\citet{hco163}]. Ideally, the
relationship between thermodynamics and mathematics should be even
more intimate. For example, general mathematical results concerning
asymptotic stability have already been obtained for metriplectic, or
GENERIC, structures by \citet{Birteaetal07}. The criteria for
thermodynamic admissibility, in particular, the existence of a
nondecreasing entropy and a positive semidefinite friction matrix,
should coincide with the criteria for mathematically proving the
existence and uniqueness of solutions to the evolution equations.
The existence of unique solutions is what we expect from meaningful
equations. Thermodynamics shall be designed to guarantee that.

\section*{Acknowledgement}
I am grateful to Markus H\"utter for several valuable suggestions to
improve the manuscript, and I thank Patrick Ilg and Martin Kr\"oger
for pointing out a number of relevant references.


\end{document}